\documentclass[sigplan,screen,nonacm]{acmart}
\usepackage[utf8]{inputenc}
\usepackage[T1]{fontenc}
\usepackage{amsmath,amsthm}
\usepackage{booktabs}
\usepackage{graphicx}
\usepackage{subcaption}
\usepackage{microtype}
\usepackage{xcolor}
\usepackage{pgfplots}
\usepackage{pgfplotstable}
\usepackage{physics}
\usepackage{enumitem}
\usepackage{multirow}
\usepackage{xfp}
\usepackage{siunitx}
\usepackage{pifont}
\usepackage{tcolorbox}
\tcbuselibrary{breakable,skins}
\usepackage{geometry}
\usepackage{hyperref}
\usepackage{threeparttable}
\usepackage{array}

\title{Real-Time Recurrent Qubit-State Discrimination on a
Heterogeneous AI-Engine--FPGA Accelerator}

\author{Vu Le}
\affiliation{%
  \institution{University of Massachusetts Amherst}
  \city{Amherst}\state{MA}\country{USA}}
\additionalaffiliation{%
\institution{Lawrence Berkeley National Laboratory}
  \city{Berkeley}\state{CA}\country{USA}}
\email{vdle@umass.edu}

\author{Yilun Xu}
\affiliation{
  \institution{Lawrence Berkeley National Laboratory}
  \city{Berkeley}
  \state{CA}
  \country{USA}
}
\email{yilunxu@lbl.gov}

\author{Neel Vora}
\affiliation{
  \institution{Lawrence Berkeley National Laboratory}
  \city{Berkeley}
  \state{CA}
  \country{USA}
}
\email{nrvora@lbl.gov}
\author{Gang Huang}
\affiliation{
  \institution{Lawrence Berkeley National Laboratory}
  \city{Berkeley}
  \state{CA}
  \country{USA}
}
\email{ghuang@lbl.gov}

\author{Deepak Ganesan}
\affiliation{
  \institution{University of Massachusetts Amherst}
  \city{Amherst}
  \state{MA}
  \country{USA}
}
\email{dganesan@umass.edu}

\author{Phuc Nguyen}
\affiliation{
  \institution{University of Massachusetts Amherst}
  \city{Amherst}
  \state{MA}
  \country{USA}
}
\email{vp.nguyen@cs.umass.edu}

\hypersetup{
  pdftitle={Real-Time Recurrent Qubit-State Discrimination on a Heterogeneous AI-Engine-FPGA Accelerator},
  pdfauthor={Vu Le; Yilun Xu; Neel Vora; Gang Huang; Deepak Ganesan; Phuc Nguyen},
  colorlinks=true,
  linkcolor=colMark,
  citecolor=colMark,
  urlcolor=colMark
}
\begin{document}
\begin{abstract}
Qubit readout, or qubit state discrimination, remains one of the
slowest and most error-prone operations in superconducting quantum
processors, making fast, accurate, and hardware-efficient classification 
essential for scalable quantum computing. Conventional discriminators
integrate the I/Q record into a single point, discarding evidence a
relaxing qubit leaves behind; full-resolution feed-forward classifiers 
recover it but are too costly to deploy under this constraint.
Recurrent classifiers retain full resolution far more
cheaply, but their loop-carried hidden state 
cannot fit a single AI Engine (AIE) tile's register
capacity and spills to data memory, adding latency that grows with
sequence length.

We present a hardware-aware LSTM accelerator for an AMD Versal 
device that resolves this by partitioning at the dependency 
boundary---independent input projections on the AIE array, the 
loop-carried update in programmable logic (PL)---keeping pace 
with a 64~ns input cadence under a 1{,}000~ns budget. On the qubit 
affected by mid-readout state transitions, our design matches a 
full-resolution feed-forward network's accuracy 
(96.59\% vs.\ 96.37\%) using $403\times$ fewer parameters and
roughly $103\times$ fewer DSPs while cutting relative error
by 68.2\% over GMM,
43.6\% over  QubiCML, and 16.82\% over MF-NN.
Using 4 out of 400 AIE tiles and 17 DSP58, it
needs $2.2$--$8.2\times$ fewer DSP58 blocks than QubiCML and
MF-NN, and the AIE--PL partition cuts exposed tail latency by
60.3\% versus a pure-AIE mapping and DSP58 usage $31.1\times$
versus an all-PL implementation. 
Across eight qubits, a compact variant reaches 96.78\%
mean accuracy, the strongest hardware-deployed operating point.
Temporal resolution and recurrent-state placement must then be treated
jointly when designing low-latency quantum readout accelerators.
\end{abstract}
\maketitle
\section{Introduction}
\label{sec:intro}

Qubit readout faces a basic timing trade-off: a longer window collects more
signal but increases exposure to relaxation and can improve separation
between quantum states.
However, it also increases the probability
that an excited qubit relaxes during measurement. When this occurs, early
samples reflect the prepared $\ket{1}$ state, while later samples increasingly
resemble $\ket{0}$. Conventional classifiers integrate these samples into a
single I/Q point, discarding when the transition occurred and potentially
moving the readout toward the wrong decision
region~\cite{hazra2407benchmarking,jung2026timeseries,luchi2023enhancing}.

This creates a fundamental challenge for low-latency readout.
A shorter
window reduces exposure to relaxation but collects less signal. A longer
window collects more signal, but integration can combine evidence from
different states and make the resulting representation less reliable.
Retaining the full temporal sequence avoids this information loss, but
increases the classifier input from two integrated values to thousands of
samples. Since signal acquisition and classification must fit within the
same real-time feedback deadline, a longer readout window
simultaneously produces more data and leaves less time to process it.

We observe this behavior in readout traces from eight superconducting
qubits. For seven qubits, classifiers operating on an integrated I/Q point
perform nearly as well as classifiers that retain the complete time series.
One qubit behaves differently. On this qubit, which we refer to as Q3, a
population of prepared $\ket{1}$ shots transitions toward $\ket{0}$ during
readout. As the window grows, GMM accuracy decreases because integration
increasingly mixes the early $\ket{1}$ evidence with the later $\ket{0}$
signal. A full-resolution classifier preserves this temporal evidence and
improves accuracy from 89.27\% with GMM and 93.95\% with QubiCML to 96.37\%.

\begin{figure*}[htbp]
    \centering
    \newcommand{\plotheight}{3cm}
    \begin{subfigure}[b]{0.32\textwidth}
        \centering
        \includegraphics[height=\plotheight]{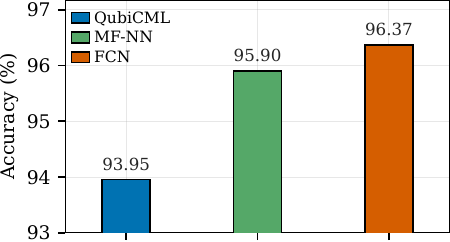}
        \caption{}
        \label{fig:baseline_acc}
    \end{subfigure}
    \hfill
    \begin{subfigure}[b]{0.32\textwidth}
        \centering
        \includegraphics[height=\plotheight]{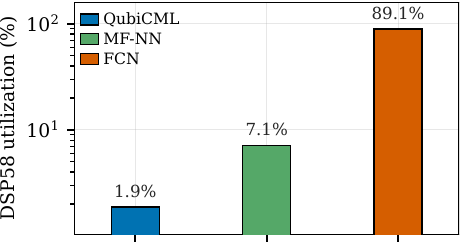}
        \caption{}
        \label{fig:baseline_dsp}
    \end{subfigure}%
    \hfill
    \begin{subfigure}[b]{0.32\textwidth}
        \centering
        \includegraphics[height=\plotheight]{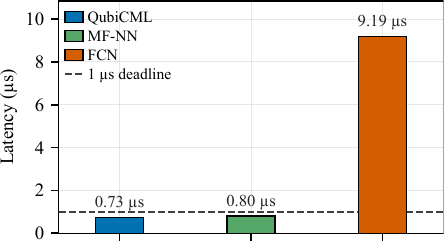}
        \caption{}
        \label{fig:baseline_latency}
    \end{subfigure}
    \caption{Comparison of single-qubit classifiers. (a) Accuracy on Q3,
    the qubit most affected by mid-readout relaxation. (b) DSP58
    utilization. (c) Latency, including the 704~ns acquisition window,
    against the 1{,}000~ns accelerator target. No existing classifier is simultaneously accurate, efficient, and fast enough. FCN values are estimates.}
    \label{fig:baseline_landscape}
\end{figure*}

Existing classifiers occupy opposite ends of the resulting
accuracy--cost tradeoff. GMM~\cite{jeffrey2014fast} and
QubiCML~\cite{vora2026ml} are inexpensive because they consume only one
integrated I/Q point, but they discard temporal information that matters on
Q3. A full-resolution FCN~\cite{lienhard2022deep} recovers this information
by consuming all 2{,}816 real values in a 704~ns trace, but requires millions
of parameters and a significant fraction of the FPGA's resources, making it impractical to deploy (Figure~\ref{fig:baseline_landscape}). The
central question is therefore: how can we retain temporal information when it
matters while meeting both a sub-microsecond latency target and a practical
per-channel hardware budget?

\paragraph{Recurrent inference}
Recurrent models offer a promising middle ground. An LSTM processes the
readout incrementally and summarizes the samples observed so far in a
fixed-size hidden state~\cite{hochreiter1997lstm,jung2026timeseries}. It can
therefore retain temporal structure without requiring a feed-forward network
whose parameter count grows with the complete input record. In our case, the
resulting full-resolution LSTM has $403\times$ fewer parameters than the FCN
while matching its accuracy on Q3. Recurrence also lets classification 
overlap with acquisition instead of waiting for the complete readout record
to arrive.

However, a compact model is not necessarily a fast or inexpensive hardware
implementation. Every LSTM timestep depends on the state produced by the
previous timestep, and the state update must complete before the next segment
of the readout arrives. At our operating point, the 2~GSPS converter produces
one 256-value input segment every 64~ns. A pure-AIE implementation cannot
sustain this rate because the loop-carried state exceeds the available
register capacity and spills to local memory. An all-PL implementation avoids
this scheduling problem, but consumes 26.9\% of the device's DSP58 budget for a single qubit, making it unscalable.
Thus, the model is compact, but it cannot be implemented efficiently on either AIE or PL alone.

\paragraph{Our approach}
Our key insight is that both temporal resolution and hardware placement
should be chosen where they provide value. At the algorithm
level, we use a full-resolution recurrent model for a qubit such as Q3,
where readout timing  affects the decision. For qubits that
do not benefit from the complete trace, we use a compact recurrent model that
preserves coarse temporal structure while reducing each timestep to one mean
I/Q pair. This avoids imposing the cost of Q3's classifier on every channel.

At the hardware level, we partition the LSTM according to its data
dependencies. The four input projections are independent and execute in
parallel on AIE tiles, while the loop-carried recurrent update and classifier
remain in PL. This keeps the hidden state local and allows each state update
to complete within the 64~ns input cadence. The AIE computation and PL
recurrent update overlap as samples arrive, leaving only the final pipeline
tail exposed after acquisition.

\paragraph{Technical challenges}
Realizing this approach requires addressing two challenges. First, selecting
a recurrent model cannot be based on accuracy alone. Hidden size, temporal
resolution, and depth jointly determine the AIE computation, recurrent update
rate, resource use, and end-to-end latency. We develop analytical feasibility
bounds that capture these constraints and eliminate configurations unable to keep pace with the input stream or meet the 1{,}000~ns latency target.
This search identifies two $H=8$ operating points: \textsf{LSTM-heavy},
which retains the complete readout record, and \textsf{LSTM-compact}, which
uses one mean I/Q pair per timestep.

Second, AIE offload is beneficial only when the input projection is large
enough to justify the AIE--PL crossing. For \textsf{LSTM-heavy}, moving the
wide input projection to the AIE array substantially reduces PL resource use
and produces a deployable design. For \textsf{LSTM-compact}, the input
projection is small, and PL-only execution is faster. Hardware placement must
therefore be selected together with the temporal representation rather than
applied uniformly across models.

\paragraph{Results}
On Q3, \textsf{LSTM-heavy} achieves 96.59\% accuracy at a 704~ns readout
duration, corresponding to a 68.2\% relative reduction in error over GMM and
43.6\% over QubiCML, and 16.82\% over MF-NN~\cite{mude2025efficient}. 
It matches the full-resolution FCN's accuracy with
$403\times$ fewer parameters and uses $2.2\times$, $3.9\times$, and
$8.2\times$ fewer DSP58 blocks than QubiCML, a prior LSTM, and MF-NN,
respectively. The AIE--PL implementation completes in 854~ns, compared with
1{,}082~ns for the pure-AIE mapping, and reduces the exposed inference tail
by 60.3\%. Relative to an all-PL implementation of the same model, it reduces
DSP58 use by $31.1\times$ and eliminates BRAM use.

The same mapping is not optimal for every model. The PL-only
\textsf{LSTM-compact} implementation completes in 814~ns, compared with
864~ns for its AIE--PL implementation, and achieves the highest mean accuracy
across the eight qubits at 96.78\%. A per-channel deployment uses
\textsf{LSTM-heavy} for Q3 and \textsf{LSTM-compact} for the other seven
qubits, outperforming the FCN in readout fidelity on every qubit. These
results show that temporal resolution and hardware placement are both
per-model design decisions.

\paragraph{Contributions}
The main contributions of this paper are:
\begin{itemize}
\item We characterize readout traces from eight superconducting qubits and
show that the value of temporal resolution is channel-dependent. In
particular, we identify a qubit for which mid-readout relaxation causes
integration to discard evidence that remains recoverable in the
time-resolved record.

\item We develop a hardware-aware recurrent design-space exploration that
jointly considers temporal resolution, hidden size, model depth, input-stream
cadence, and end-to-end latency. The resulting compact and full-resolution
operating points allow classifier complexity to follow the needs of each
qubit.

\item We develop a heterogeneous recurrent implementation that maps
independent LSTM input projections to AIE tiles and keeps the loop-carried
state update in PL. This dependency-aware partition sustains the input cadence
while substantially reducing PL resource use.

\item We implement and evaluate the design on an AMD Versal VCK190, achieving
FCN-level accuracy on the qubit most affected by relaxation within an
854~ns accelerator latency, while a per-channel deployment provides high
readout fidelity across all eight qubits. We confirm this latency on physical
VCK190 hardware for the deployed \textsf{LSTM-heavy} AIE--PL mapping,
measuring 860~ns end-to-end.
\end{itemize}

\section{Background}
\label{sec:background}
\subsection{Superconducting Qubit Readout}

\begin{figure}[htbp]
    \centering
    \includegraphics[width=\linewidth]{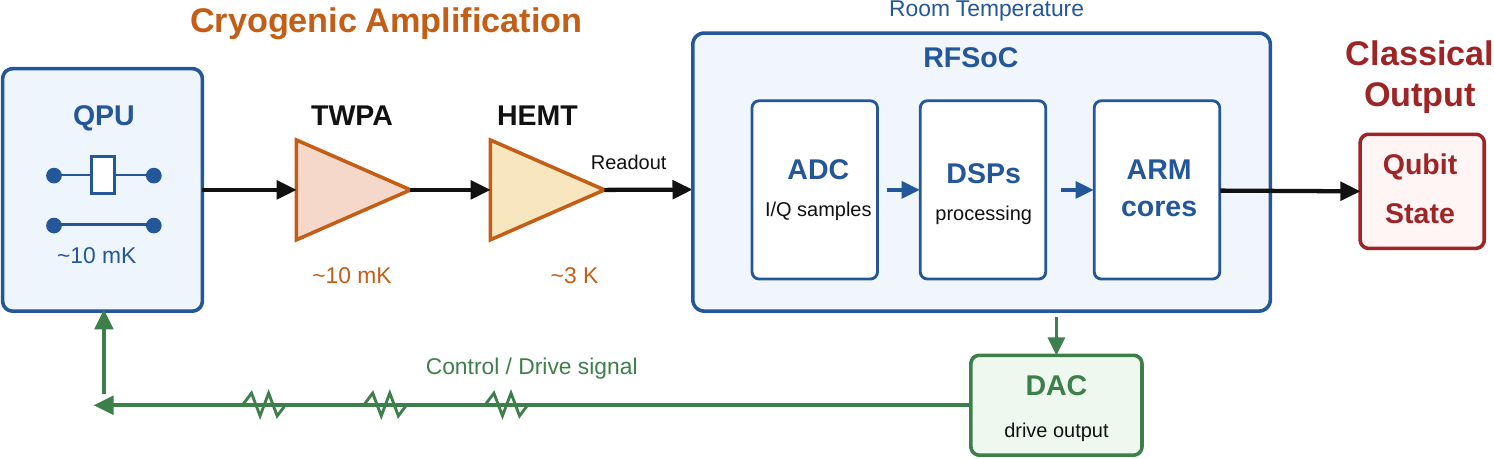}
 	\caption{Schematic of a typical superconducting transmon qubit control and readout chain.}
    \label{fig:control_system}
\end{figure}

Figure~\ref{fig:control_system} shows a typical readout pipeline in
superconducting quantum platforms. Qubit state information is extracted via
dispersive measurement: a microwave probe pulse traverses a resonator
coupled to the qubit, acquiring state-dependent phase and amplitude shifts
that are digitized into in-phase (I) and quadrature (Q) components
\cite{xu2023qubic,stefanazzi2022qick}. The resulting I/Q samples form a
time series whose trajectory captures the temporal evolution of the readout
signal, including transient state changes that may occur during
measurement.

On the QubiC platform~\cite{xu2023qubic}, a 2~GSPS sampling rate and finite
measurement duration together produce a large volume of time-series data
per shot. These traces are typically processed with matched filters or
machine-learning classifiers on the control board or an external host, and
this step is often the bottleneck: it is slow relative to the qubit
coherence time and prone to misclassification. Readout accuracy depends
on both the analog front end and this discriminator stage; this paper
focuses on the latter, where the accuracy--latency trade-off we study is
decided.

\subsection{Temporal Readout Classification}

Integration-based classifiers commonly reduce a readout record to a single
point in the IQ plane and apply a statistical or feed-forward discriminator.
Gaussian-mixture models~\cite{jeffrey2014fast} are attractive because they are simple and inexpensive,
while feed-forward neural networks can learn more expressive decision
boundaries. Both approaches, however, operate on the representation presented
to them and cannot recover temporal information removed by integration.

A sequence-aware classifier instead processes the readout as a temporal
trajectory. LSTMs \cite{hochreiter1997lstm} maintain a recurrent hidden state
across timesteps and can therefore use the evolution of the measurement record
to distinguish trajectories that have similar integrated IQ values. Prior work
has demonstrated LSTM-based reconstruction of superconducting-qubit trajectories
\cite{koolstra2022monitoring} and sequence-based qubit-state discrimination
\cite{jung2026timeseries}. These results motivate temporal inference for
readout, but do not imply that the full raw trajectory is necessary for
every workload: temporal aggregation can trade input resolution for
substantially lower model, hardware, and communication cost.
Section~\ref{sec:workload} characterizes this trade-off across an
eight-qubit workload before the hardware design is introduced.
\section{The Case for Time-Resolved Readout}
\label{sec:workload}



% ...








\begin{figure*}[htbp]
    \centering
    \includegraphics[width=\linewidth]{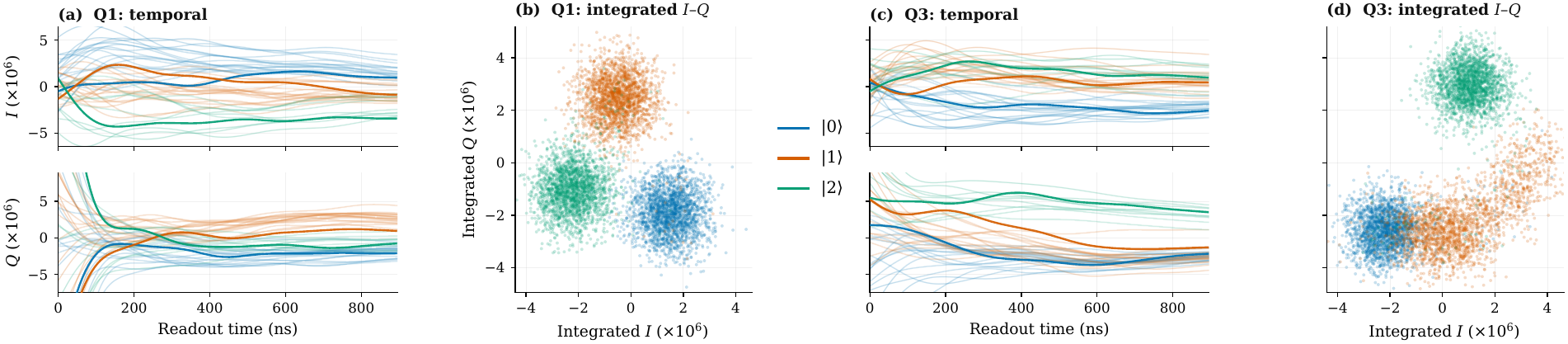}
    \caption{Readout trajectories for a representative well-behaved qubit (Q1) and the integration-limited qubit (Q3). (a),(c) Running I/Q trajectories; (b),(d) the corresponding integrated points. Q1 remains separable after integration. On Q3, a population of prepared $\ket{1}$ shots transitions toward $\ket{0}$ during readout, causing the integrated distributions to overlap. \emph{Traces are bandpass-filtered for visual clarity only.}}
    \label{fig:q1q3_traj}
\end{figure*}


The core observation behind our design is that a single input representation does not serve all qubits. 
For a qubit such as Q1, conventional integration can reduce the full readout trace to two accumulated I/Q values while preserving the information needed for classification.
For a qubit such as Q3, however, integration obscures when the state changes during measurement. Its classifier must retain more of the temporal record and consequently process a larger input. Thus, the temporal resolution required for qubit-state discrimination, and the hardware resources needed to process that input, are channel-dependent.

Figure~\ref{fig:q1q3_traj} contrasts these two regimes. On Q1, the three prepared states remain well separated after integration, and retaining the complete temporal record provides little additional value. On Q3, a population of prepared $\ket{1}$ shots moves toward the $\ket{0}$ trajectory during readout, causing their integrated points to overlap with the $\ket{0}$ cluster. The early samples still contain evidence of the prepared state, but that evidence is lost when the complete record is reduced to one integrated I/Q point.

This comparison raises two questions. First, how much of the accuracy loss can be recovered by learning a better decision boundary over the same integrated input? Second, how much additional accuracy requires retaining temporal information?

We answer these questions by comparing classifiers along the input-resolution axis, from one integrated I/Q point to the complete readout trace. These comparisons make the case for selecting temporal resolution per channel and show how that choice changes the hardware workload. Section~\ref{sec:eval_accuracy} later compares recurrent and feed-forward models after controlling the input representation more closely.

\subsection{Experimental Setup and Comparison}

We evaluate single-shot readout traces from eight fixed-frequency transmon qubits at
Lawrence Berkeley National Laboratory, acquired with QubiC~\cite{xu2023qubic}. The dataset
contains 25{,}000 shots per prepared state $\ket{0}$, $\ket{1}$, and 
$\ket{2}$. The ADC samples at 2~GSPS, so the 704~ns window used throughout the 
evaluation contains $S=1408$ complex samples, or $2S=2816$ interleaved I/Q 
values. The classifier consumes unfiltered post-DLO I/Q samples rescaled to 
the int16 range. We discard shots that fail heralded-state 
preparation~\cite{johnson2012heralded} ($\sim$1--7\%, by qubit) and report
accuracy on the retained shots.

We use three classifiers to expose the accuracy--resolution tradeoff. GMM~\cite{jeffrey2014fast} and QubiCML~\cite{vora2026ml} consume the same integrated I/Q point, or two real values, but use different decision functions. Their comparison shows what can be gained by learning a more expressive boundary without retaining additional temporal information. The FCN~\cite{lienhard2022deep} consumes the complete 2{,}816-value trace. We use it as a full-resolution accuracy reference, while noting that its millions of parameters and high hardware cost make it impractical for our target deployment. All classifiers use the training procedure described in Section~\ref{sec:eval_setup}.


\subsection{Impact of Temporal Resolution}
\label{sec:workload-across}

\begin{figure}[htbp]
    \centering
    \includegraphics[width=\linewidth]{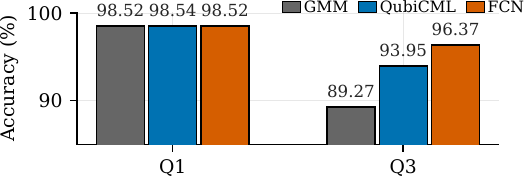}
    \caption{Readout accuracy on Q1 and Q3 at 704~ns. Retaining the complete temporal record provides little benefit on Q1 but substantially improves accuracy on Q3.}
    \label{fig:q1_q3_acc_bar_classifiers}
\end{figure}

The accuracy results show the same two regimes (Figure~\ref{fig:q1_q3_acc_bar_classifiers}). On Q1, classifier choice has little effect because integration preserves a separable representation. Q3 shows a different progression. At 704~ns, GMM reaches 89.27\% accuracy. QubiCML improves this to 93.95\% using a learned boundary over the same two integrated values. Thus, 4.68  points of the gap can be recovered without changing the input representation. The full-resolution FCN reference reaches 96.37\%, indicating that another 2.42 points are available when the temporal record is retained. This gap between the integrated QubiCML and the full-resolution FCN shows that learning a better decision boundary alone is insufficient on Q3.

The readout-duration sweep provides a stronger test of this result. If integration only needed more signal, its accuracy should improve or eventually saturate as the window grows. On Q3, extending the window instead makes the integrated representation less reliable.

\begin{figure}[htbp]
    \centering
    \includegraphics[width=\linewidth]{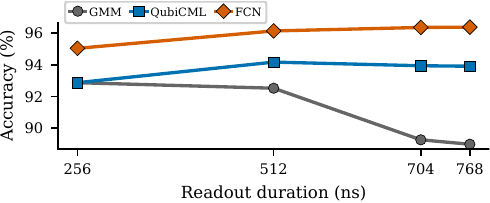}
    \caption{Accuracy versus readout duration on Q3. GMM declines as the window grows, QubiCML plateaus after 512~ns, and the full-resolution FCN continues to improve.}
    \label{fig:q3_acc_vs_duration}
\end{figure}

As shown in Figure~\ref{fig:q3_acc_vs_duration}, GMM accuracy is nearly flat through 512~ns at 92.53\% and then falls to 88.99\% by 768~ns. The longer window makes the integrated statistic less discriminative because it mixes more of the later $\ket{0}$-like signal into relaxing $\ket{1}$ shots. QubiCML recovers a better boundary in the integrated plane but plateaus once that representation has discarded the transition timing. The FCN continues to improve because its input retains both the early $\ket{1}$ evidence and the subsequent trajectory. Together with Figure~\ref{fig:q1q3_traj}, this shows that the problem is not simply insufficient signal or an inadequate decision boundary; on Q3, useful information is lost during integration.


Across the eight qubits, Q3 has the largest measured benefit from retaining the complete trace. The other seven channels do not exhibit the same continuing gain at full resolution (Table~\ref{tab:acc_8q_704ns}). The channels therefore require unequal temporal resolution: Q3 benefits from the complete trace, whereas the other seven can use a more compact representation.

\subsection{Implications for Hardware--Algorithm Co-Design}
\label{sec:workload-implications}

The input representation should therefore be selected per channel rather than fixed across the device. A difficult channel such as Q3 should retain the complete temporal record, while other channels can use a compact representation without paying the cost of processing the full trace.

This choice directly changes the hardware workload: retaining more temporal detail increases the input dimension and the cost of processing each timestep. Hence, we need a hardware-aware approach that selects how much temporal information each channel retains while ensuring that the resulting classifier can sustain the input cadence and meet the latency and resource budgets.

\section{Hardware-Aware Design-Space Exploration}
\label{sec:design}

Building on the observations in Section~\ref{sec:workload}, our approach has two main components: selecting the temporal representation and recurrent-model configuration to match each channel, and selecting the hardware placement for the resulting workload. Both decisions must account for the input cadence, accelerator latency, and available hardware resources. This section develops the analytical model and design-space exploration used to make these choices.

\subsection{Target Platform and Execution Model}
\label{sec:target_platform}

We explore these design choices for the AMD Versal VCK190, whose combination of AI Engines (AIEs) and programmable logic (PL) exposes the mapping tradeoff central to our approach. The device provides a 400-tile AIE array at 1.25~GHz connected to PL through PLIO/AXI4-Stream interfaces~\cite{ds950}. Each AIE tile provides a 512-bit
vector datapath and 32~KB of local SRAM, supporting up to 32 MACs per
cycle for int16 operands, while the PL fabric provides customizable datapaths.
This asymmetry motivates the mapping used throughout the paper: the wide,
independent $W_x\mathbf{x}_t$ projections are mapped to AIE, while the small
loop-carried recurrent update is implemented in PL.
These platform characteristics determine where the LSTM computation can execute. Whether a mapping is feasible, however, also depends on how quickly readout samples arrive and when the classification result must be returned.

\subsection{Streaming Interface and Design Objective}
\label{sec:design_objective}

\begin{figure}[htbp]
    \centering
    \includegraphics[width=0.95\linewidth]{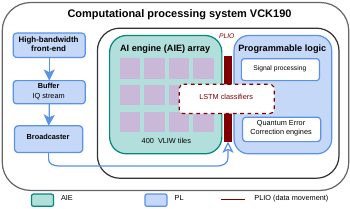}
    \caption{Target streaming architecture for real-time quantum-state
    classification on the AMD Versal VCK190. I/Q samples from the QubiC
    ZCU216 RFSoC enter the PL domain and are streamed to the AIE input
    projection engines through PLIO.}
    \Description{Block diagram of the streaming architecture for real-time quantum-state classification.}
    \label{fig:system_overview}
\end{figure}

We show the target system in Figure~\ref{fig:system_overview}. The quantum control system QubiC~\cite{xu2023qubic} digitizes the readout signal at
2~GSPS and streams complex I/Q samples toward the VCK190 via the high-bandwidth
front-end, where PL forwards them to the AIE input-projection engines through PLIO 
as they arrive, without buffering the complete readout burst.

We report \emph{accelerator latency} from the first I/Q sample leaving the
converter interface to assertion of the valid class label. This excludes the
analog front end, digital down-conversion, inter-board transfer, and controller
return path, none of which is implemented in our setup. 
We use the 1{,}000~ns target as an admission bound for the accelerator and
reproduce the 2~GSPS input cadence offline with a paced replay buffer;
no real-time cold-qubit experiments were performed.

Future RF-series Versal devices~\cite{versalrf} integrate RF-ADCs/DACs with the
same PL/AIE fabric, eliminating the inter-board communication our reported
latency already excludes.

\subsection{Recurrent Workload Decomposition}
\label{sec:lstm_structure}
To determine how an LSTM maps to this streaming system, we first separate the computation that depends on the current input from the loop-carried state update. The LSTM forward pass~\cite{hochreiter1997lstm,gers2000learning} at timestep $t$ is
\begin{equation}
\setlength{\arraycolsep}{1pt}
\begin{gathered}
\begin{bmatrix}\mathbf{i}_t & \mathbf{f}_t & \mathbf{g}_t & \mathbf{o}_t\end{bmatrix}^{\!\top}
=
\begin{bmatrix}\sigma & \sigma & \tanh & \sigma\end{bmatrix}^{\!\top}
\!\left(W_x\mathbf{x}_t+W_h\mathbf{h}_{t-1}+\mathbf{b}\right),\\[4pt]
\mathbf{c}_t=\mathbf{f}_t\odot\mathbf{c}_{t-1}
+\mathbf{i}_t\odot\mathbf{g}_t,\qquad
\mathbf{h}_t=\mathbf{o}_t\odot\tanh(\mathbf{c}_t).
\end{gathered}
\label{eq:lstm}
\end{equation}

Here $\mathbf{i}_t$, $\mathbf{f}_t$, $\mathbf{g}_t$, and $\mathbf{o}_t$ are the input, forget, candidate, and output gates, respectively, while $W_x$ and $W_h$ stack their input and recurrent weight matrices. The equation exposes two computational terms central to our mapping. The input projection $W_x\mathbf{x}_t$ depends only on the current input, so the four gate-specific projections can execute in parallel as $\mathbf{x}_t$ arrives. The recurrent projection $W_h\mathbf{h}_{t-1}$ depends on the previous hidden state and therefore lies on the loop-carried dependency path.

For input dimension $X$ and hidden dimension $H$, each gate's input projection requires $HX$ MACs per timestep, while its recurrent projection requires $H^2$. At the $H=8$ operating point evaluated later, a raw 64~ns segment has $X=256$, making the input projection $32\times$ larger than the recurrent projection. Reducing the same segment to one mean I/Q pair gives $X=2$, making the recurrent projection four times larger than the input projection. The temporal representation therefore changes not only the input size, but also where the LSTM spends its work.

To simplify the loop-carried state update, we replace the standard sigmoid and tanh with piecewise-linear hardsigmoid and hardtanh nonlinearities that can be implemented as clamps:
\begin{equation}
\begin{gathered}
\sigma_h(x) = \operatorname{clamp}\!\left(\frac{x}{4}+\frac{1}{2},0,1\right),\\
\tanh_h(x) = \operatorname{clamp}(x,-1,1).
\end{gathered}
\label{eq:hard_activations}
\end{equation}

We also clamp the cell state to $[-1,1]$, so the hidden-state update reduces to $\mathbf{h}_t=\mathbf{o}_t\odot\mathbf{c}_t$.

\subsection{Analytical Hardware Model}
\label{sec:aie_model}

The decomposition above allows us to determine whether a candidate LSTM can keep pace with the input stream. We model the AIE input-projection time, the PL recurrent-update time, and the exposed pipeline tail, then use these terms to test both streaming feasibility and the 1{,}000~ns accelerator target. We use the following platform constants:
\begin{equation}
\begin{aligned}
f_{\mathrm{AIE}} &= 1.25~\mathrm{GHz}, &
\mathrm{MAC}_{\mathrm{peak}} &= 32, \\
B_{\mathrm{AIE}} &= 64~\text{bit/cycle}.
\end{aligned}
\label{eq:platform_constants}
\end{equation}

The 32-MAC/cycle peak follows from the 512-bit AIE vector datapath, while
$B_{\mathrm{AIE}}$ follows from the two 32-bit input-stream ports~\cite{ds950,ug1079}.
The nominal weight-load bandwidth is therefore eight times the 
activation-stream bandwidth.

\subsubsection{Input-Projection Lower Bound}
\label{sec:lemma1}

For one gate tile computing $\mathbf{z}_k=W_{ik}\mathbf{x}_t$, two independent
constraints bound the projection time: compute requires at least
$HX/\mathrm{MAC}_{\mathrm{peak}}$ AIE cycles, and input delivery requires at
least $X/B_{\mathrm{eff}}$ cycles, where
$B_{\mathrm{eff}}=\min(B_{\mathrm{AIE}},B_{\mathrm{link}})$ is the effective
activation bandwidth in int16 values per AIE cycle, capped by either the
native AIE path ($B_{\mathrm{AIE}}=4$ int16/cycle) or a lower upstream link bandwidth limited by clock-domain crossing (CDC), $B_{\mathrm{link}}$. Thus
\begin{equation}
    T_{\mathrm{Wx}}^{\mathrm{floor}}(H,X)
      = \max\!\left(
          \frac{HX}{\mathrm{MAC}_{\mathrm{peak}}},
          \frac{X}{B_{\mathrm{eff}}}
        \right).
    \label{eq:gate_floor}
\end{equation}
Because the four gate projections execute independently and in parallel, this
single-gate bound also determines the AIE input projection cadence floor; it
captures only fundamental compute and input-delivery costs, so implementation
overheads can only increase the realized cadence.

The native AIE balance point is
$H^*_{\mathrm{AIE}}=\mathrm{MAC}_{\mathrm{peak}}/B_{\mathrm{AIE}}=32/4=8$:
sustaining 32 int16 MACs/cycle requires 512~bits/cycle of weights, against a
native input path of 64~bits/cycle. The system-level balance is
$H^*_{\mathrm{sys}}=\mathrm{MAC}_{\mathrm{peak}}/B_{\mathrm{eff}}$. For
$H<H^*_{\mathrm{AIE}}$, a single gate projection is input-stream bound; in our
mapping, $H=4$ packs two independent gate projections per tile, using two AIE
tiles for all four gates. For $H>H^*_{\mathrm{AIE}}$, a single tile becomes
compute-bound at the native interface; $H=16$ therefore requires two tiles per
gate to keep pace with the system stream. Thus $H^*_{\mathrm{AIE}}=8$ is an
intrinsic tile property, while the realized system balance depends on the
effective upstream bandwidth $B_{\mathrm{eff}}$.

\subsubsection{Streaming Feasibility}
\label{sec:stream_feasibility}
Let $S$ denote the number of complex I/Q samples in the selected readout
window and $T$ the number of LSTM timesteps, so each timestep represents
$B=S/T$ complex I/Q samples. For an ADC sampling at $f_{\mathrm{ADC}}$, a
timestep therefore arrives every $T_{\mathrm{bin}}=B/f_{\mathrm{ADC}}$; for
our workload, $f_{\mathrm{ADC}}=2$~GSPS, so $T_{\mathrm{bin}}=0.5B$~ns.

The AIE input projection is hidden by acquisition, and the PL recurrent
update keeps pace with the stream, when
\begin{equation}
T_{\mathrm{AIE}}\,T_{\mathrm{Wx}}^{\mathrm{floor}}(H,X)\leq T_{\mathrm{bin}}
\quad\text{and}\quad
T_{\mathrm{rec}}^{\mathrm{step}}(H,N_L)\leq T_{\mathrm{bin}},
\label{eq:stream_conditions}
\end{equation}
where $T_{\mathrm{AIE}}=0.8$~ns is the AIE clock period on the VCK190, $N_L$ is the number of LSTM layers and $T_{\mathrm{rec}}^{\mathrm{step}}$
is the PL state-update time at the target 500~MHz clock.

When both stages sustain the input cadence, only the exposed tail is added to
the acquisition time:
\begin{equation}
    T_{\mathrm{acc}}^{\mathrm{est}}(H,T,N_L)
    = \frac{S}{f_{\mathrm{ADC}}}
      + T_{\mathrm{tail}}(H,T,N_L)
    \leq 1{,}000~\mathrm{ns}.
    \label{eq:e2e_est}
\end{equation}
For the target architecture,
\begin{equation}
\begin{aligned}
T_{\mathrm{tail}}(H,T,N_L)
  ={}& T_{\mathrm{plio}}
     +T_{\mathrm{rec}}^{\mathrm{tail}}(H,T,N_L)\\
    & +(N_L-1)T_{\mathrm{il}}
     +T_{\mathrm{dense}}(N_L)
     +T_{\mathrm{CDC}}^{\mathrm{exposed}},
\end{aligned}
\label{eq:tail_decomp}
\end{equation}
where $T_{\mathrm{CDC}}^{\mathrm{exposed}}$ includes only CDC latency that
remains exposed after the streaming stages have completed. For the deployed
$H=8$, one-layer datapath, the measured recurrent update is
$T_{\mathrm{rec}}^{\mathrm{step}}=31$ PL cycles $=62$~ns, of which 13 cycles
are overlapped with the input stream,
$T_{\mathrm{plio}}=6$~ns, and $T_{\mathrm{dense}}=36$~ns. The
$1{,}000$~ns bound is an admission test: a candidate that exceeds it cannot
meet the target even if excluded control-path stages consume no accelerator
budget. Candidates that pass are implemented in hardware.

\subsection{Pure-AIE Mapping Constraints}
\label{sec:failure_modes}
The input-projection balance does not by itself make the recurrent update
efficient on AIE. A pure-AIE mapping must either localize the loop-carried state
on a central tile or distribute it across tiles.
\paragraph{State localization.}
A central tile aggregates the four gate results and maintains
$\mathbf{h}_t$ and $\mathbf{c}_t$ locally. For the implemented kernel, the
architectural live state can be approximated by
\begin{equation}
    N_{\mathrm{live}}(H)=
\underbrace{2\left\lceil\frac{4H}{8}\right\rceil}_{\scriptsize\text{bias}}
+\underbrace{\left\lceil\frac{H}{4}\right\rceil}_{\mathbf{h}_t}
+\underbrace{\left\lceil\frac{H}{4}\right\rceil}_{\mathbf{c}_t}
+\underbrace{\left\lceil\frac{H}{8}\right\rceil}_{\mathbf{h}_{16}}
+\underbrace{2}_{\scriptsize\text{wt.\ temp.}}.
\end{equation}

Counts are in 128-bit vector registers, of which the AIE tile provides
16~\cite{ug1079}. At $H=8$, the architectural state alone is approximated at
13 registers, with the weight temporaries adding 2 more for a predicted
total of 15; the compiler assigns 14 of the 16
physical registers, closely matching this estimate. The pure-AIE kernel
consequently spills to data memory: the compiler reports
\texttt{\_\_spill\_DM\_stack} and seven register-allocation moves, and the
measured II is 100 cycles versus a 64-cycle floor. At 1.25~GHz, this is
80~ns, exceeding the 64~ns input cadence. At $H=4$, the same architectural
state occupies only 7 registers.
\paragraph{State distribution.}
Distributing the recurrent update across AIE tiles reduces local register
pressure but requires the previous hidden state to cross the interconnect at
every timestep, making gate and state update kernels as well as scheduling
a lot more complex. The communication therefore recurs $T$ times for only $H^2$
recurrent MACs per gate per timestep. We did not implement this variant, so
the comparison is analytical rather than measured. Pure-AIE mapping thus
trades register pressure for repeated state movement, motivating placement of
the loop-carried state in PL while the independent input projections remain on
AIE tiles.

\subsection{Hardware-Constrained Design-Space Exploration}
\label{sec:nas}

\begin{figure}[htbp]
    \centering
    \includegraphics[width=1\linewidth]{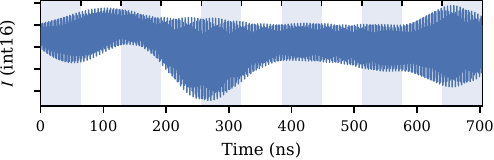}
    \caption{Raw Q3 trace ($I$ component, int16-rescaled), with alternating
shading marking successive $T{=}11$, $B{=}128$-sample (64~ns)
\textsf{LSTM-heavy} timesteps. }
    \label{fig:lstm_heavy_window}
\end{figure}

The preceding analysis yields three feasibility checks: the AIE input projection
and PL recurrent update must each keep pace with the input stream, and the
acquisition-plus-tail latency must remain within the 1{,}000~ns budget. We use
these analytical tests to prune candidate models before training.
We explore a refined search region over hidden dimension, temporal resolution,
representation, and depth:
\begin{itemize}[leftmargin=*, noitemsep]
\item Hidden dimension: $H\in\{4,8,16\}$.
\item Temporal bin size: $B\in\{128,256,512\}$ complex I/Q samples per timestep.
\item Temporal representation:
      $\mathcal{R}\in\{\mathrm{raw},\mathrm{meanIQ}\}$, giving
      $X=2B$ (raw) and $X=2$ (meanIQ), respectively.
\item Number of LSTM layers: $N_L\in\{1,2\}$.
\end{itemize}
Recall $S$, the number of complex I/Q samples in the selected readout window
(Sec.~\ref{sec:stream_feasibility}); the number of complete timesteps is
$T=\lfloor S/B\rfloor$, with leftover samples discarded.
For the primary 704~ns workload, the 2~GSPS input
gives $S=1408$ complex samples. The deployed configurations use $B=128$,
corresponding to a 64~ns timestep and $T=11$. \textsf{LSTM-heavy} therefore
consumes $X=256$ real interleaved I/Q values per timestep, whereas
\textsf{LSTM-compact} reduces the same 64~ns window to one mean I/Q pair,
giving $X=2$ while keeping the same 11-step sequence.
Figure~\ref{fig:lstm_heavy_window} visualizes this binning on a raw Q3 trace
($I$ only; $Q$ is binned identically).

We additionally verify that the resident gate weights fit within the 31~KB
of tile-local memory available after reserving 1~KB for the stack: for a
tile processing $H_{\mathrm{tile}}$ output rows, the resident weight
footprint is $M_{\mathrm{gate}}(H_{\mathrm{tile}},X,d)=H_{\mathrm{tile}}Xd
\leq31{,}744$ bytes, where $d=2$~bytes for the int16 weights used by the
deployed models. All candidates in this refined search region satisfy this
capacity bound, so local memory is a residency check rather than a binding
pruning constraint.

The joint admission rule combining these four checks is therefore
\begin{equation}
\begin{aligned}
\operatorname{admit}(H,\mathcal{R},T,N_L)\iff\;&
T_{\mathrm{AIE}}\,
T_{\mathrm{Wx}}^{\mathrm{floor}}(H,X)
\leq T_{\mathrm{bin}}\\
&\wedge\;
T_{\mathrm{rec}}^{\mathrm{step}}(H,N_L)\leq T_{\mathrm{bin}}\\
&\wedge\;
T_{\mathrm{acc}}^{\mathrm{est}}(H,T,N_L)\leq1{,}000~\mathrm{ns}\\
&\wedge\;
M_{\mathrm{gate}}(H_{\mathrm{tile}},X,d)\leq31{,}744~\mathrm{bytes}
\end{aligned}
\label{eq:nas_joint}
\end{equation}
where $X=X(\mathcal{R},B)$ and $H_{\mathrm{tile}}$ is determined by the
gate-to-tile partitioning used for the candidate.

The resulting search exposes three regimes. With $B=64$, the PL recurrent
update is the limiting stage for $H=8$ and cannot sustain the input cadence.
At the deployed $B=128$ point, both the AIE projection and recurrent update
keep pace. For larger bins, the AIE input projection becomes the limiting
stage even relative to the 100-cycle recurrent II of the spilled pure-AIE
kernel; these candidates remain stream-feasible but provide lower accuracy
than $B=128$.

\subsubsection{Search Outcome}
\label{sec:nas_search_result}

The search identifies single-layer $H{=}8$ as the best balance between
accuracy and hardware cost. A larger hidden dimension ($H{=}16$) doubles
the AIE tiles required for input projection without improving Q3
accuracy; coarser temporal binning ($B{=}256,512$) reduces accuracy
while lowering projection cost; and a second recurrent layer remains
cadence-feasible but adds a second recurrent engine and inter-layer
relay cost with no accuracy benefit. Full ablation numbers are given in
Appendix~\ref{sec:hardware_aware_search_outcome}. We therefore deploy
single-layer $H{=}8$ designs: \textsf{LSTM-compact} ($X{=}2$) and
\textsf{LSTM-heavy} ($X{=}256$).

\section{Co-Designed LSTM Implementation}
\label{sec:implementation}

The hardware-aware design-space exploration identifies two operating
points with $H=8$: \textsf{LSTM-heavy}, which preserves the raw I/Q samples
within each timestep, and \textsf{LSTM-compact}, which reduces each 64~ns
timestep to one mean I/Q pair. Both use the same heterogeneous AIE--PL design;
only the input-projection kernel changes.

The implementation partitions the LSTM at its data dependency boundary. The
four independent input projections $\mathbf{z}_k=W_{ik}\mathbf{x}_t$ execute
on four AIE tiles, while the loop-carried recurrent update and the final
classifier remain in programmable logic. 

\begin{figure}[htbp]
    \centering
    \includegraphics[width=1\linewidth]{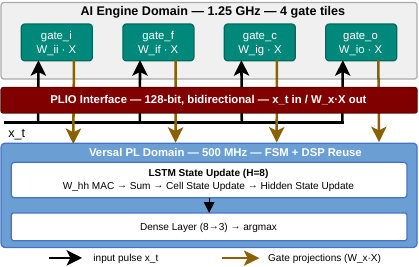}
	\caption{Architecture of the AIE--PL LSTM. Four AIE tiles compute the
	input projections $W_{ik}\mathbf{x}_t$ in parallel; a PL state-update engine
	produces the three-class output.}
    \label{fig:lstm_arch}
\end{figure}

\subsection{AIE Input-Projection Engine}
\label{sec:impl_gate}

Each projection $\mathbf{z}_k=W_{ik}\mathbf{x}_t$ ($k\in\{i,f,g,o\}$) runs
on its own AIE tile. For \textsf{LSTM-heavy}, $H=8$ and $X=256$ real interleaved
int16 values are consumed per timestep. The analytical lower bound is 64 AIE
cycles; the implemented kernel achieves 73 cycles, or 58.4~ns at 1.25~GHz,
when driven at the native dual-stream peak of 4~int16 values/AIE cycle.
The additional 9 cycles capture input delivery, data type
conversion, and PLIO output setup omitted from the lower-bound model.

The deployed 2~GSPS input arrives at 64~Gb/s, corresponding to one 256-value
timestep every 64~ns. The 73-cycle kernel interval is therefore below the
64~ns arrival interval, so the input projection is stream-paced rather than
compute-paced. FIFOs absorb the rate mismatch between the peak AIE
throughput and the converter stream.

\textsf{LSTM-compact}'s AIE--PL mapping shares this datapath and recurrent
backend, but its input-projection kernel consumes one mean I/Q pair
($X=2$) per 64~ns timestep. A single 32-bit PLIO stream 
is sufficient for each tile.

\subsection{PL Recurrent-State Engine}
\label{sec:impl_pl}
Gate results cross the PLIO streaming interface, are buffered across the
clock-domain crossing, and are reassembled by a PL-side merge stage before
the state update. The recurrent update and dense classifier are 
implemented as a finite-state
controller coupled to a time-multiplexed DSP58 MAC datapath in 
PL at 500~MHz. The engine computes
$W_{hk}\mathbf{h}_{t-1}$, updates $\mathbf{c}_t$ and $\mathbf{h}_t$, and
the dense classifier head, while reusing DSP58 resources across
these operations. Hardsigmoid and hardtanh are implemented in LUT logic.

At $H=8$, the
recurrent projection takes 13 PL cycles (26~ns), followed by 18 cycles
(36~ns) for the state update, giving a 31-cycle (62~ns) steady-state update.
The recurrent projection overlaps the streamed AIE input projection; hence the
18-cycle portion is the exposed PL critical path.

\subsection{Streaming Overlap and Critical Path}
\label{sec:impl_timing}
\begin{figure}[htbp]
    \centering
    \includegraphics[width=\linewidth]{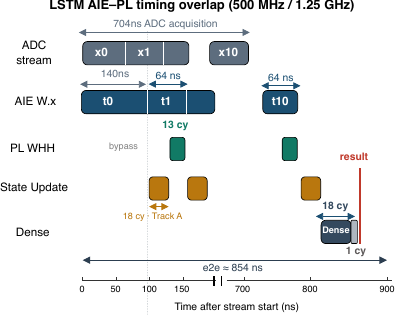}
    \caption{Streaming datapath overlap for the deployed $H=8$, $T=11$
    implementation. AIE input projection overlaps PL recurrent processing at
    the 64~ns input cadence; the 31-cycle PL update has an 18-cycle exposed
    state-update portion.}
    \label{fig:timing_breakdown_aie_pl_h8}
\end{figure}

AIE projection and PL recurrent processing execute concurrently as input
streams through the pipeline (Fig.~\ref{fig:timing_breakdown_aie_pl_h8}), so
only the 18-cycle PL update (Section~\ref{sec:impl_pl}) is exposed at the
state-update boundary for the deployed $H{=}8$, $T{=}11$ design.


Hardware-integration simulation reproduces the 64~ns gate cadence for timesteps
1--11. The final input leaves an exposed drain before the last gate result
reaches PL; this drain and the final PL state-update/output work form the
latency tail. The deployed \textsf{LSTM-heavy} implementation therefore has
704~ns of acquisition plus a 150~ns exposed tail, for the 854~ns accelerator
latency reported in Section~\ref{sec:evaluation}. The corresponding pure-AIE
implementation exposes a 378~ns tail.

The hybrid pipeline is deterministic: all datapaths are statically scheduled
and contain no data-dependent control flow, so repeated shots produce the same
accelerator latency.

\subsection{Pure-AIE Baseline}
\label{sec:impl_pure_aie}

For a fair comparison with the AIE-PL version, we instantiate a pure-AIE $H=8$
design using the same four input-projection tiles and a central recurrent-state
tile with the same hardtanh and hardsigmoid. The register-pressure-induced 100-cycle initiation interval is analyzed and measured in Section~\ref{sec:failure_modes};
its end-to-end latency is reported in Section~\ref{sec:evaluation}. 
No PL resources are attributed to this baseline.

\section{Evaluation}
\label{sec:evaluation}

We evaluate our hardware-aware LSTM design across
classification quality, hardware cost, and real-time latency. We
focus on four questions: (1) does the selected recurrent model retain the
temporal-information benefit on mid-readout state transition
workloads; (2) how does its accuracy compare with the 
feed-forward and other machine learning-based classifiers;
(3) what resource cost does the heterogeneous AIE--PL implementation require; and
(4) how does the measured execution compare with the pure-AIE baseline and
the sub-microsecond latency objective?

\subsection{Experimental Setup}
\label{sec:eval_setup}

\textbf{Model training.}
All learned models are implemented in PyTorch~\cite{paszke2019pytorch} and
trained for 100 epochs with Adam optimizer~\cite{kingma2014adam}, an initial learning rate
of $10^{-4}$, batch size 128, and exponential decay every
10 epochs. All classifiers use the same stratified 80/20 split and
random seed. 
Our LSTMs use the hardware-friendly hardsigmoid and hardtanh
nonlinearities described in Section~\ref{sec:lstm_structure} for both training and inference.
Because the prior LSTM~\cite{jung2026timeseries} has no hardware implementation,
we synthesize its architecture with our own hard activations, trained with its reported
GMM-confidence-weighted loss, for the resource and latency figures in Section~\ref{sec:eval_latency}; accuracy figures instead use its original
soft activations. It also does not specify a bin width, so we use 64~ns
bins for fair comparison with our recurrent models.

\textbf{Hardware evaluation.}
We use \texttt{aiesimulator}~\cite{ug1076} for AIE kernel timing and functional behavior,
Verilator~\cite{verilator} for the PL datapath and AIE--PL integration checks, and QEMU~\cite{bellard2005qemu} with the \texttt{v++}~\cite{ug1076} system link
for end-to-end system execution and latency measurement.
Synthesis and implementation use Vivado~2025.2 at a 500~MHz
constraint for a fair latency comparison; resource utilization 
and WNS come from
post-implementation reports, with post-synthesis figures flagged where used.
The PL logic and all LSTM variants, including the prior
LSTM~\cite{jung2026timeseries}, are written in Chisel~\cite{bachrach2012chisel}
for one parameterized design; QubiCML~\cite{vora2026ml} and
MF-NN~\cite{mude2025efficient}, 
which reduces each trace to nine matched-filter likelihood scores
before a small feed-forward network refines the classification,
are implemented with hls4ml~\cite{fahim2021hls4ml} and hls~\cite{UG1399}. 
All baselines are our own re-implementations at their published 
configurations, thus differences from published numbers are possible.

\subsection{Impact on Readout Accuracy}
\label{sec:eval_accuracy}

\begin{figure*}[htbp]
    \centering
    \includegraphics[width=\linewidth]{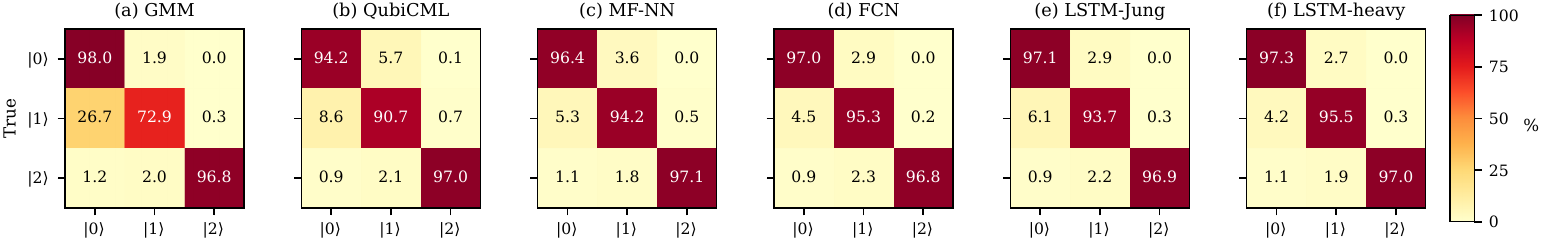}
   \caption{Row-normalized confusion matrices for Q3 at 704\,ns. Cell values
show the percentage of samples in each true class assigned to each predicted
class. The dominant GMM error is $|1\rangle\!\rightarrow\!|0\rangle$
misclassification, which is substantially reduced by the full-resolution FCN
and recurrent models.}
    \label{fig:q3_confusion_matrix}
\end{figure*}

\begin{table}[htbp]
\centering
\caption{Single-shot readout accuracy (\%) at 704\,ns ($T{=}11$, FP32
evaluation). GMM\textsuperscript{a}~\cite{jeffrey2014fast},
QubiCML~\cite{vora2026ml}, MF-NN~\cite{mude2025efficient},
FCN\textsuperscript{a}~\cite{lienhard2022deep}, LSTM~\cite{jung2026timeseries};
LSTM-h/LSTM-c denote LSTM-heavy/LSTM-compact.
\textsuperscript{a}accuracy-only references, not synthesized. Full results
with binomial standard errors are provided in Appendix~\ref{sec:appendix_windows}.}
\label{tab:acc_8q_704ns}
\setlength{\tabcolsep}{2.5pt}
\resizebox{\columnwidth}{!}{%
\begin{tabular}{lccccccc}
\toprule
Qubit & GMM & QubiCML & MF-NN & FCN & LSTM & LSTM-h & LSTM-c \\
\midrule
Q0 & 97.70 & 97.77 & 98.00 & 98.08 & \textbf{98.18} & 97.96 & 98.15 \\
Q1 & 98.52 & 98.54 & 98.52 & 98.52 & \textbf{98.75} & 98.26 & 98.74 \\
Q2 & 97.52 & 97.67 & 97.67 & 97.59 & 97.88 & 97.41 & \textbf{97.89} \\
Q3 & 89.27 & 93.95 & 95.90 & 96.37 & 95.87 & \textbf{96.59} & 96.11 \\
Q4 & 94.72 & 94.39 & 95.59 & 95.22 & 95.67 & 94.96 & \textbf{95.74} \\
Q5 & 97.85 & 97.89 & 97.81 & 97.98 & 98.20 & 97.93 & \textbf{98.25} \\
Q6 & 93.01 & 93.62 & 92.68 & 93.75 & 94.15 & 93.92 & \textbf{94.27} \\
Q7 & 94.71 & 94.70 & 94.63 & 94.91 & \textbf{95.15} & 94.79 & 95.10 \\
\midrule
$\mu \pm \sigma$ & 95.41 & 96.07 & 96.35 & 96.55 & 96.73 & 96.48 & \textbf{96.78} \\
\bottomrule
\end{tabular}%
}
\end{table}
Q3 is the qubit most affected by mid-readout relaxation
(Section~\ref{sec:workload}), so we use it as our central 
comparison case. GMM and QubiCML, operating on an integrated
point, reach 89.27\% and 93.95\%, respectively, while every
structure-preserving model exceeds 95.8\%; LSTM-heavy is highest at
96.59\%, a 68.2\% relative reduction in error over GMM, 43.6\% over
QubiCML, 16.82\% over MF-NN and 6.06\% over the FCN, though  
not significant after correction ($p=0.06$). 


Figure~\ref{fig:q3_confusion_matrix} shows why Q3 differs:
GMM misassigns roughly a quarter of prepared $\ket{1}$ shots to $\ket{0}$,
an error every structure-preserving model cuts to about 4\%, while the
$\ket{2}$ row remains largely unchanged. The recovered population is 
consistent with the mid-readout relaxation events  in
Section~\ref{sec:workload}.

Across eight qubits, LSTM-compact achieves the highest mean accuracy
(96.78\%), a 29.9\% relative reduction in error over GMM and 6.7\% over the
closest baseline, FCN, with QubiCML and MF-NN in between (18.1\% and
11.8\%), despite far fewer parameters than the FCN.

In addition to pooled accuracy, we  report readout fidelity,
a stricter, class-balanced metric: the geometric mean of per-state classification
accuracy, $F = (P(0|0)\cdot P(1|1)\cdot P(2|2))^{1/3}$, computed from the
row-normalized confusion matrix on the same frozen eval split.
A per-channel deployment (\textsf{LSTM-compact} everywhere except Q3, using \textsf{LSTM-heavy}) outperforms the FCN in readout fidelity on every qubit
(Table~\ref{tab:deployed_vs_fcn}), by 0.07 to 0.54 points (0.29 on average).

\begin{table}[htbp]
\centering
\caption{Deployed per-channel readout fidelity (\%) vs.\
the FCN reference~\cite{lienhard2022deep},
at 704~ns.
Deployed uses \textsf{LSTM-compact} on all channels except Q3, where
\textsf{LSTM-heavy} is used.}
\label{tab:deployed_vs_fcn}
\setlength{\tabcolsep}{3pt}
\resizebox{\columnwidth}{!}{%
\begin{tabular}{lcccccccc}
\toprule
 & Q0 & Q1 & Q2 & Q3 & Q4 & Q5 & Q6 & Q7 \\
\midrule
FCN$^{*}$ & 98.07 & 98.52 & 97.59 & 96.36 & 95.19 & 97.96 & 93.64 & 94.83 \\
Deployed & \textbf{98.14} & \textbf{98.74} & \textbf{97.89} & \textbf{96.59} & \textbf{95.71} & \textbf{98.24} & \textbf{94.19} & \textbf{95.01} \\
\bottomrule
\end{tabular}%
}
\end{table}

\subsection{Q3 Accuracy--Resource Comparison}
\label{sec:eval_q3_resources}
\begin{figure}[htbp]
    \centering
    \includegraphics[width=1\linewidth]{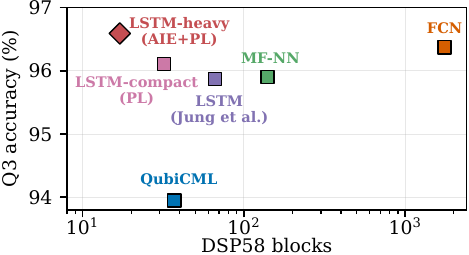}
    \caption{Accuracy on Q3 against DSP58 usage (log scale) for each
    classifier's best deployable mapping.}
    \label{fig:q3_accuracy_vs_dsp}
\end{figure}

Figure~\ref{fig:q3_accuracy_vs_dsp} compares accuracy against DSP58
usage on Q3.
DSP58 is the binding resource for every classifier evaluated, 
so this comparison determines hardware feasibility rather
than latency alone. 
The resource gap is largest here: at 96.59\% on Q3, \textsf{LSTM-heavy} needs
$3.9\times$ fewer DSP58 blocks than the prior LSTM (95.87\%), $8.2\times$
fewer than MF-NN (95.90\%), and $2.2\times$ fewer than QubiCML.
\textsf{LSTM-compact} trails
\textsf{LSTM-heavy} by only 0.48 points (96.11\%) while using the fewest
DSP58 blocks of any classifier evaluated, including QubiCML.
\textsf{LSTM-heavy} matches FCN's 96.37\%;
\textsf{LSTM-compact} comes within 0.26 points,
at a fraction of FCN's estimated 89.1\% DSP58 budget,
consistent with prior
reports that this architecture scales poorly on FPGA fabric
\cite{maurya2023scaling};
full synthesis exhausted our host
memory during elaboration, so we report an hls4ml-based analytical
estimate using a per-layer $\mathrm{ReuseFactor}$ equal to each
layer's fan-in (2816/1000/500/250) instead; latency is estimated the
same way, extrapolated from smaller synthesized networks to
9.19~\textmu s.

DSP58 is not the only axis worth checking: Table~\ref{tab:pl_resources}
reports LUT and BRAM usage alongside DSP58 and AI Engine tile count for
every design. \textsf{LSTM-heavy}'s DSP58 and BRAM advantage
does not extend to LUT, where it uses more than either baseline; we report the
full resource profile rather than DSP58 alone, since 
Section~\ref{sec:eval_resources} details how that DSP58 advantage is achieved.


\begin{table}[htbp]
\centering
\caption{Resource usage on the VCK190 for different classifiers in this paper.}
\label{tab:pl_resources}
\begin{tabular}{lrrrr}
\toprule
Design & DSP58 & LUT & BRAM & AIE tiles \\
\midrule
QubiCML       & 37 (1.9\%)   & 0.18\% & 0.0\% & --- \\
MF-NN         & 140 (7.1\%)  & 0.42\% & 7.5\% & --- \\
FCN & 1{,}753 (89.1\%) & ---    & ---   & --- \\
LSTM-compact  & 32 (1.6\%)   & 0.52\% & 0.8\% & 0/400 \\
LSTM-heavy    & 17 (0.86\%)  & 0.58\% & 0.0\% & 4/400 \\
\bottomrule
\end{tabular}
\end{table}


\subsection{PL vs. AIE+PL Resource Comparison}
\label{sec:eval_resources}

\begin{figure}[htbp]
    \centering
    \includegraphics[width=1\linewidth]{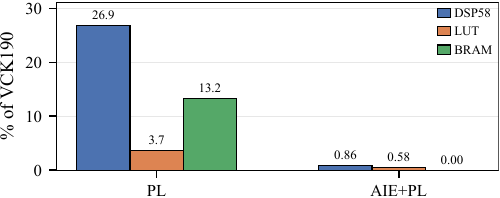}
    \caption{LSTM-heavy resource use on the VCK190, as a fraction of
    device DSP58 and LUT capacity, for the PL-only and the AIE+PL
    implementation. 
    }
    \label{fig:heavy-pl-vs-aie-pct}
\end{figure}

The AIE--PL \textsf{LSTM-heavy} mapping offloads its wide $8\times256$ input
projection to 4 out of 400 AIE tiles, cutting DSP58 usage from 26.9\% to 0.86\% of the
device and LUT usage from 3.7\% to 0.58\%
(Figure~\ref{fig:heavy-pl-vs-aie-pct}), a $31.1\times$ reduction in DSP58
blocks and $6.4\times$ in LUTs, while eliminating all 128 BRAMs.

This is what lets the AIE offload bring a full-resolution recurrent model
under budget on this device (Section~\ref{sec:eval_q3_resources}).
\textsf{LSTM-compact} already fits in PL at 32 DSP58 blocks, so we
retain its PL-only mapping rather than offloading to AIE: with only
$X{=}2$ input features per timestep, the projection is too small for
the AIE round trip to pay for itself.

\subsection{End-to-End Latency}
\label{sec:eval_latency}
\begin{figure}[htbp]
    \centering
    \includegraphics[width=1\linewidth]{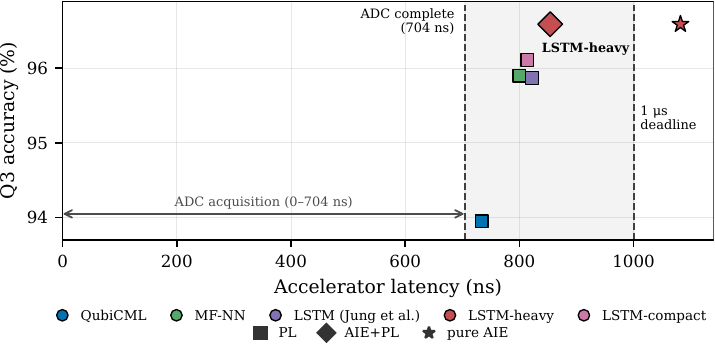}
    \caption{Measured accelerator latency under a 704-ns acquisition stream.}
    \label{fig:latency_vs_accuracy}
\end{figure}

Figure~\ref{fig:latency_vs_accuracy} plots accelerator latency against mean
accuracy for every evaluated classifier and mapping, relative to the 704~ns
acquisition window and the 1{,}000~ns deadline. Table~\ref{tab:e2e_latency}
compares the evaluated classifiers under the same 2~GSPS replay conditions
(accelerator latency as defined in Section~\ref{sec:design_objective}). As 
detailed in Section~\ref{sec:impl_timing}, the AIE--PL
and pure-AIE mappings reach the
same 96.59\% accuracy on Q3 since both implement identical trained weights,
but only the AIE--PL mapping's shorter exposed tail meets the 1{,}000~ns
deadline. On physical VCK190 hardware, this AIE--PL mapping measures
860~ns end-to-end for a representative single shot---within 0.7\% of the
854~ns simulated figure above.

\begin{table}[htbp]
\centering
\caption{Single-shot accelerator latency at the 704\,ns operating point.
Tail is accelerator latency minus the acquisition window; $\Delta$ is the
tail reduction of each model's deployable mapping over its alternative.
$^{\dagger}$Measured on physical VCK190 hardware: 860~ns.
$^{\ddagger}$Reference only.}
\label{tab:e2e_latency}
\small
\setlength{\tabcolsep}{5pt}
\resizebox{\columnwidth}{!}{%
\begin{tabular}{llrrr}
\toprule
Model & Mapping & Accel. (ns) & Tail (ns) & $\Delta$ (\%) \\
\midrule
QubiCML~\cite{vora2026ml}      & PL & 734 & 30  & -- \\
MF-NN~\cite{mude2025efficient} & PL & 800 & 96  & -- \\
LSTM~\cite{jung2026timeseries} & PL & 822 & 118 & -- \\
\midrule
\multicolumn{5}{l}{\emph{This work}} \\
\textit{LSTM-compact}
& \textbf{PL only} & \textbf{814} & \textbf{110} &
\multirow{2}{*}{$31.25$} \\
& AIE+PL & 864 & 160 & \\
\addlinespace[2pt]
\textit{LSTM-heavy}
& \textbf{AIE+PL} & \textbf{854}$^{\dagger}$ & \textbf{150} &
\multirow{2}{*}{$60.32$} \\
& Pure AIE & 1082 & 378 & \\
\cmidrule(l){2-5}
& PL only$^{\ddagger}$ & 792 & 88 & -- \\
\bottomrule
\end{tabular}%
}
\end{table}

An all-PL \textsf{LSTM-heavy} implementation is nominally faster at
792~ns but misses timing closure at 500~MHz by a narrow margin
($\mathrm{WNS}=-0.005$~ns); we use it only as the all-PL
resource-comparison reference, not as a deployable operating point. For
\textsf{LSTM-compact}, PL-only completes in
814~ns versus 864~ns for AIE+PL: with only $X{=}2$ input features per timestep, the AIE offload
saves too little projection work to amortize the fixed AIE--PL crossing.

\section{Related Work}
\label{sec:related}
\textbf{Feed-forward inference for qubit readout.}
Vora et al.~\cite{vora2026ml} demonstrated low-latency
FPGA inference for qubit/qutrit discrimination---the QubiCML
baseline we reimplement and evaluate in
Section~\ref{sec:eval_q3_resources}---while Butko et
al.~\cite{butko2026qsd} extended feed-forward inference to the Versal
AI Engine. Maurya et al.~\cite{maurya2023scaling} reduced classifier
complexity through matched filters and a small neural network to fit
FPGA resource constraints, and Mude et al.~\cite{mude2025efficient}
extended this approach to three-level readout. Concurrent work by
Otting et al.~\cite{otting2026multi} targets the same failure mode with
a Mamba-based sequence model, reducing parameter count by 49.6\% over
the prior state of the art, but evaluates entirely in software. None of
these establish a hardware mapping for a recurrent classifier's
loop-carried state; Published latencies across this body of work are
also not directly comparable, since acquisition windows differ,
so we reimplement selected baselines under the same
704~ns replay and measurement definition.

\textbf{Recurrent inference on spatial accelerators.}
Prior AIE mappings such as CHARM~\cite{zhuang2023charm} and
MaxEVA~\cite{taka2023maxeva} optimize dense matrix
multiplication for throughput, whereas Sapkas et al.~\cite{sapkas2025gru}
study latency-constrained GRU inference on Versal and keep
recurrent matrix-vector operations AIE-resident, proposing a
AIE--PL variant for faster result aggregation. Our partition is
complementary: the independent input projections execute on AIE tiles,
while the loop-carried recurrent state remains in PL---the reverse of
their assignment.

\textbf{Hardware-aware recurrent model selection.} 
Hardware-aware NAS methods such as ProxylessNAS\cite{cai2019proxylessnas} 
and MnasNet\cite{tan2019mnasnet} use learned or measured cost models to 
navigate large architecture spaces. Recurrent FPGA accelerators such as
ESE\cite{han2017ese} and ML-to-FPGA toolkits like 
hls4ml\cite{duarte2018fast,fahim2021hls4ml} instead target fixed 
architectures, trading throughput for resource use through compression
or reuse-factor scheduling.
\section{Conclusion}
\label{sec:conclusion}

In this paper, we show that recurrent inference resists spatial
acceleration at small hidden sizes because register pressure, not
arithmetic throughput, becomes the bottleneck---our $H{=}8$ pure-AIE
kernel spills to data memory and settles at a 100-cycle initiation
interval against a 64-cycle floor---and we resolve this by placing the
loop-carried state in PL and the four independent input projections on
AIE tiles. This partition cuts \textsf{LSTM-heavy}'s exposed tail from
378 to 150~ns and its accelerator latency from 1{,}082 to 854~ns (860~ns
measured on VCK190), inside the 1{,}000~ns budget, while reducing PL
DSP58 usage by 96.8\% and LUT usage by 84.3\% and eliminating BRAM use,
using 4/400 AI Engine tiles. Against prior baselines,
\textsf{LSTM-heavy} uses $2.2$--$8.2\times$ fewer DSP58 blocks than
QubiCML, the prior LSTM, and MF-NN, and matches a full-resolution FCN's
accuracy on Q3 (96.59\% vs.\ 96.37\%) with roughly $403\times$ fewer
parameters; \textsf{LSTM-compact} outperforms QubiCML and MF-NN in
accuracy on every qubit. Results are otherwise obtained offline via
simulation, using a paced replay buffer, with no real-time cold-qubit
experiments; integrating the accelerator into a live control loop and
extending this partition to multiple simultaneous channels, where AIE
tiles and PLIO interfaces become shared resources, are natural next
steps.

\begin{acks}
This work was supported in part by the National Science
Foundation (NSF) under Grant Nos. 2403528 and 2449995;
the U.S. Department of Energy (DOE), Office of Science,
through the Quantum Systems Accelerator, a National
Quantum Information Science Research Center (Award No.
DESCL0000121); and the Laboratory Directed Research and
Development (LDRD) Program at Lawrence Berkeley National
Laboratory under DOE Contract No. DE-AC02-05CH11231.

We thank Mark Rollins, Jose Manuel Monsalve Diaz, and
Paul Hartke of Advanced Micro Devices, Inc. (AMD) for
helpful technical discussions and assistance with
implementation and debugging.

Yilun Xu has a financial interest in OpenQI, Inc.

Claude (Anthropic) was used to assist with editing and
revising portions of this manuscript. The authors are
responsible for the final content.
\end{acks}

\bibliographystyle{plain}
\bibliography{refs}
\appendix
\section{Hardware-Aware Search Outcome}
\label{sec:hardware_aware_search_outcome}

Table~\ref{tab:nas_results} shows representative outcomes
at the 704\,ns operating point: $H{=}16$ ties $H{=}8$ at 96.59\% while
doubling AIE tiles to eight; wider timesteps ($B{=}256,512$, which
repartition the raw sequence into fewer, larger steps without discarding
samples) lower projection cost but cost 0.04--0.17 accuracy points; a
second recurrent layer adds 576 parameters for no accuracy gain (96.58\%
vs.\ 96.59\%); and a narrower timestep ($B{=}64$) is rejected outright,
as the PL recurrent update cannot keep pace with the resulting cadence.
We therefore deploy only the two highlighted configurations.

\begin{table}[htbp]
\centering
\caption{Representative outcomes of the hardware feasibility tests at the
704\,ns operating point. $B$ denotes complex I/Q samples per timestep and $X$
denotes real interleaved I/Q values presented to the LSTM at each timestep.}
\label{tab:nas_results}
\small
\setlength{\tabcolsep}{4pt}
\renewcommand{\arraystretch}{1.05}
\begin{tabular}{@{}cccccc
        >{\raggedright\arraybackslash}p{0.42\columnwidth}@{}}
\toprule
$H$ & $B$ & $X$ & $T$ & $N_L$ & Q3 Acc. & Outcome \\
\midrule
\textbf{8}  & \textbf{128} & \textbf{2}   & \textbf{11} & \textbf{1} & 96.11\% &
\textsf{LSTM-compact} \\
\textbf{8}  & \textbf{128} & \textbf{256} & \textbf{11} & \textbf{1} & 96.59\% &
\textsf{LSTM-heavy} \\
\midrule
8  & 256 & 512  & 5 & 1 &
96.55\% & admitted; not deployed \\
8  & 512 & 1024 & 2 & 1 &
96.42\% & admitted; not deployed \\
\midrule
8  & 128 & 256  & 11 & 2 &
96.58\% & admitted; not deployed \\
16 & 128 & 256  & 11 & 1 &
96.59\% & admitted with 8 AIE tiles; not deployed \\
8  & 64  & 128  & 22 & 1 &
-- & rejected: PL recurrent cadence \\
\bottomrule
\end{tabular}
\end{table}

\section{Ablation Study}

\subsection{Impact of Input Quantization}
\label{sec:input_quantization}
The results throughout this paper (e.g., Table~\ref{tab:acc_8q_704ns})
evaluate FP32 models on I/Q samples already rescaled to the int16 range
(Section~\ref{sec:workload}), matching QubiC's native fixed-point ADC
output. Unlike Section~\ref{sec:quantization}, which quantizes the model
itself, here the model stays FP32 and only the input representation
changes. Table~\ref{tab:int16_vs_fp32} compares FP32 and int16 inputs for
\textsf{LSTM-heavy} and a separately trained FCN reference.

\begin{table}[htbp]
\centering
\caption{Single-shot accuracy (\%) at 704\,ns (raw). $\Delta$ = int16 $-$ fp32 (percentage points). Mean is across 8 qubits.}
\label{tab:int16_vs_fp32}
\small
\begin{tabular}{@{}lrrrrrr@{}}
\toprule
qubit & \multicolumn{3}{c}{LSTM-heavy} & \multicolumn{3}{c}{FCN} \\
\cmidrule(lr){2-4}\cmidrule(lr){5-7}
 & fp32 & int16 & $\Delta$ & fp32 & int16 & $\Delta$ \\
\midrule
Q0 & 97.97 & 97.97 & +0.00 & 98.10 & 98.09 & -0.01 \\
Q1 & 98.26 & 98.25 & -0.01 & 98.51 & 98.52 & +0.01 \\
Q2 & 97.41 & 97.31 & -0.10 & 97.61 & 97.61 & +0.00 \\
Q3 & 96.59 & 96.60 & +0.01 & 96.35 & 96.36 & +0.01 \\
Q4 & 94.97 & 95.07 & +0.10 & 95.31 & 95.31 & +0.00 \\
Q5 & 97.95 & 97.75 & -0.20 & 97.97 & 97.98 & +0.01 \\
Q6 & 93.91 & 93.94 & +0.03 & 93.77 & 93.78 & +0.01 \\
Q7 & 94.83 & 94.68 & -0.15 & 94.91 & 94.91 & -0.01 \\
\midrule
$\mu$ & 96.49 & 96.45 & -0.04 & 96.57 & 96.57 & +0.00 \\
\bottomrule
\end{tabular}
\end{table}


The per-qubit accuracy change stays within 0.20 percentage points for
either model, and the mean shifts by at most 0.04 points; minor
differences from Table~\ref{tab:acc_8q_704ns}'s numbers for either model
reflect ordinary training variance across separate runs, not the input
change under test.

\subsection{Hard Activation versus Soft Activation}
\label{sec:hard_soft_activation}
The mean difference is $+0.03$ points for LSTM-heavy and $-0.04$ for
LSTM-compact, an order of magnitude below the mean accuracy differences
the paper's conclusions rest on, so the hardware-friendly activations are
free in accuracy terms on average. Q4 is the largest single-qubit
exception ($+0.29$ for LSTM-heavy).
The LSTM-heavy/LSTM-compact comparisons elsewhere in this paper are
therefore unaffected by activation choice, since both models are trained
and evaluated with the same hard activations throughout.

\begin{table}[htbp]
\centering
\caption{Activation ablation at $704\,\mathrm{ns}$ ($T{=}11$).
Hard activations (piecewise-linear sigmoid and
$\mathrm{hardtanh}$) versus true $\sigma$ and $\tanh$. $\Delta$ is soft minus
hard (percentage points).}
\label{tab:ablation-activations}
\begin{tabular}{l ccc ccc}
\toprule
 & \multicolumn{3}{c}{LSTM-heavy} & \multicolumn{3}{c}{LSTM-compact} \\
\cmidrule(lr){2-4} \cmidrule(lr){5-7}
Qubit & Hard & Soft & $\Delta$ & Hard & Soft & $\Delta$ \\
\midrule
Q0 & 97.96 & 97.98 & $+0.02$ & 98.15 & 98.16 & $+0.01$ \\
Q1 & 98.26 & 98.31 & $+0.05$ & 98.74 & 98.71 & $-0.03$ \\
Q2 & 97.41 & 97.48 & $+0.07$ & 97.89 & 97.83 & $-0.06$ \\
Q3 & 96.59 & 96.52 & $-0.07$ & 96.11 & 96.06 & $-0.05$ \\
Q4 & 94.96 & 95.25 & $+0.29$ & 95.74 & 95.68 & $-0.06$ \\
Q5 & 97.93 & 97.92 & $-0.01$ & 98.25 & 98.20 & $-0.05$ \\
Q6 & 93.92 & 93.94 & $+0.02$ & 94.27 & 94.21 & $-0.06$ \\
Q7 & 94.79 & 94.68 & $-0.11$ & 95.10 & 95.09 & $-0.01$ \\
\midrule
\textbf{Mean} & \textbf{96.48} & \textbf{96.51} & $\mathbf{+0.03}$ & \textbf{96.78} & \textbf{96.74} & $\mathbf{-0.04}$ \\
\bottomrule
\end{tabular}
\end{table}

\subsection{Impact of Quantization}
\label{sec:quantization}
The accuracy results above use FP32 evaluation of models trained with the
hardware-faithful hardsigmoid and hardtanh functions. The deployed
datapath instead uses int16 arithmetic (Q13 weights, Q15 activations); we
quantize only our deployed models, since baselines elsewhere in this
paper are not part of our hardware deployment.
Table~\ref{tab:quant} compares int16 accuracy against FP32 on the same
eight-qubit test sets.

\begin{table}[htbp]
\centering
\caption{Float32 versus deployed int16 (Q13 weights, Q15 activations) accuracy (\%).}
\label{tab:quant}
\small
\setlength{\tabcolsep}{5pt}
\begin{tabular}{lrrr rrr}
\toprule
& \multicolumn{3}{c}{LSTM-heavy} & \multicolumn{3}{c}{LSTM-compact} \\
\cmidrule(lr){2-4} \cmidrule(lr){5-7}
Qubit & FP32 & int16 & $\Delta$ & FP32 & int16 & $\Delta$ \\
\midrule
Q0 & 97.96 & 97.95 & $-0.01$ & 98.15 & 98.14 & $-0.01$ \\
Q1 & 98.26 & 98.10 & $-0.16$ & 98.74 & 98.73 & $-0.01$ \\
Q2 & 97.41 & 97.41 & $+0.00$ & 97.89 & 97.90 & $+0.01$ \\
Q3 & 96.59 & 96.57 & $-0.02$ & 96.11 & 96.05 & $-0.06$ \\
Q4 & 94.96 & 94.80 & $-0.16$ & 95.74 & 95.70 & $-0.04$ \\
Q5 & 97.93 & 97.85 & $-0.08$ & 98.25 & 98.24 & $-0.01$ \\
Q6 & 93.92 & 93.90 & $-0.02$ & 94.27 & 94.27 & $+0.00$ \\
Q7 & 94.79 & 94.86 & $+0.07$ & 95.10 & 95.11 & $+0.01$ \\
\midrule
\textbf{Mean} & \textbf{96.48} & \textbf{96.43} & $\mathbf{-0.05}$ & \textbf{96.78} & \textbf{96.77} & $\mathbf{-0.01}$ \\
\bottomrule
\end{tabular}
\end{table}

Quantization has negligible impact: mean accuracy drops by at most 0.05
points, and the largest single-qubit loss is 0.16 points (Q1, Q4,
LSTM-heavy). The temporal-resolution and Q3 findings reported above
therefore hold for the deployed int16 representation.

\section{Bandwidth--Accuracy Tradeoff: Temporal Compression on Q3}
\label{sec:eval_binning}

\begin{figure}[htbp]
    \centering
    \includegraphics[width=\linewidth]{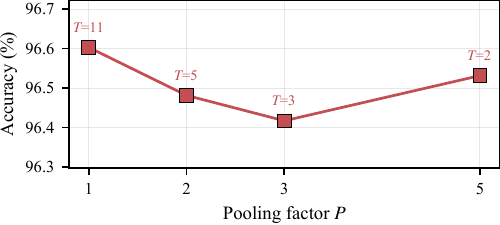}
    \caption{LSTM-heavy accuracy on Q3 vs.\ pooling factor $B$ on the
    fixed 128-sample timestep, at 704~ns. Labels report the resulting
    number of timesteps $T$.}
    \label{fig:binning_study}
\end{figure}

Reducing per-shot data volume allows more qubit streams to share a fixed
interboard bandwidth. We test this by pooling the raw Q3 I/Q stream by a
factor $P$ before it reaches LSTM-heavy's fixed 128-feature timestep---unlike
the timestep-width sweep in Section~\ref{sec:nas_search_result}, pooling
here averages raw samples together, so information is destroyed rather
than reorganized. As $P$ grows, fewer timesteps are needed to cover the
704~ns window ($T{=}11$ at $P{=}1$ down to $T{=}2$ at $P{=}5$), and
accuracy stays within 0.18 points of deployed throughout
(Figure~\ref{fig:binning_study}). This robustness makes LSTM-heavy
suitable for Q3, where raw resolution matters most
(Section~\ref{sec:eval_q3_resources}); a contrasting failure mode for
LSTM-compact is given in Appendix~\ref{sec:appendix_binning}.

\section{Temporal Binning on LSTM-compact}
\label{sec:appendix_binning}
\begin{figure}[htbp]
    \centering
    \includegraphics[width=\linewidth]{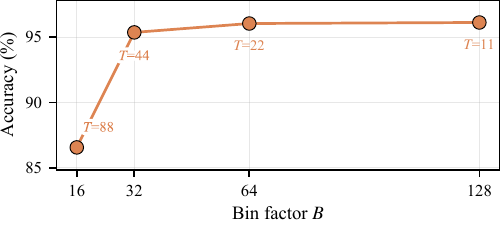}
    \caption{LSTM-compact accuracy on Q3 vs.\ mean-IQ averaging window
    $B$, at 704~ns. Labels report the resulting number of timesteps $T$.}
    \label{fig:binning_compact_appendix}
\end{figure}

Section~\ref{sec:eval_binning} shows that LSTM-heavy tolerates pooling
of the raw Q3 stream because pooling only stretches timestep width,
leaving per-feature averaging fixed. LSTM-compact has no such fixed
per-feature width: each timestep reduces directly to a single mean over
$B$ raw samples, so $B$ sets the per-timestep noise floor rather than
timestep width alone. Accuracy is nearly flat through $B{=}64$ (96.03\%,
within 0.08 points of deployed), drops to 95.35\% at $B{=}32$ (0.76
points below deployed), and collapses to 86.56\% at $B{=}16$
(Figure~\ref{fig:binning_compact_appendix}), where too few raw samples
remain in each average for the added temporal resolution ($T$ rising to
88) to compensate.

\section{Accuracy Across Readout Windows}
\label{sec:appendix_windows}
Accuracy improves substantially from 256 to 512~ns and continues improving
through 704~ns, after which returns collapse: extending to 768~ns adds only
0.06 points to mean LSTM-compact accuracy (0.07 for LSTM-heavy), and on Q3,
LSTM-heavy moves from 96.59\% to 96.60\%. We adopt 704~ns as the operating
point for the hardware evaluation, since it captures essentially all of the
accuracy available from a longer window while returning 64~ns of the
1~\textmu s control budget and shortening each shot's exposure to relaxation
during measurement.

\begin{table*}[htbp]
\centering
\caption{Single-shot readout accuracy (\%) at 256\,ns. Conventions as in Table~\ref{tab:acc_8q_704ns}; $^{\ast}$accuracy-only, never synthesized.}
\label{tab:acc_8q_256ns}
\setlength{\tabcolsep}{4pt}
\begin{tabular}{lccccccc}
\toprule
Qubit & GMM~\cite{jeffrey2014fast}$^{\ast}$ & QubiCML~\cite{vora2026ml} & MF-NN~\cite{mude2025efficient} & FCN$^{\ast}$~\cite{lienhard2022deep} & LSTM~\cite{jung2026timeseries} & LSTM-heavy & LSTM-compact \\
\midrule
Q0 & 97.21$\pm$0.14 & 97.06$\pm$0.14 & 97.11$\pm$0.14 & 97.19$\pm$0.14 & \textbf{97.35$\pm$0.13} & 97.16$\pm$0.14 & 97.34$\pm$0.13 \\
Q1 & 87.78$\pm$0.27 & 87.75$\pm$0.27 & 88.40$\pm$0.26 & 88.15$\pm$0.27 & 88.46$\pm$0.26 & 87.99$\pm$0.27 & \textbf{88.59$\pm$0.26} \\
Q2 & 92.88$\pm$0.22 & 92.73$\pm$0.22 & 93.17$\pm$0.21 & 93.01$\pm$0.22 & 93.19$\pm$0.21 & 92.53$\pm$0.22 & \textbf{93.31$\pm$0.21} \\
Q3 & 92.88$\pm$0.22 & 92.88$\pm$0.22 & 93.74$\pm$0.20 & 95.05$\pm$0.18 & 93.84$\pm$0.20 & \textbf{95.29$\pm$0.18} & 93.95$\pm$0.20 \\
Q4 & 63.91$\pm$0.40 & 64.50$\pm$0.40 & 66.44$\pm$0.39 & 66.10$\pm$0.39 & 66.33$\pm$0.39 & 66.11$\pm$0.39 & \textbf{66.86$\pm$0.39} \\
Q5 & 91.41$\pm$0.23 & 91.05$\pm$0.24 & 91.84$\pm$0.23 & 91.58$\pm$0.23 & 92.03$\pm$0.22 & 91.44$\pm$0.23 & \textbf{92.13$\pm$0.22} \\
Q6 & 87.70$\pm$0.27 & 87.60$\pm$0.27 & 88.06$\pm$0.27 & 88.01$\pm$0.27 & \textbf{88.08$\pm$0.27} & 87.87$\pm$0.27 & 88.01$\pm$0.27 \\
Q7 & 94.52$\pm$0.19 & 94.44$\pm$0.19 & 94.23$\pm$0.19 & 94.48$\pm$0.19 & \textbf{94.60$\pm$0.19} & 94.25$\pm$0.19 & \textbf{94.60$\pm$0.19} \\
\midrule
$\mu \pm \sigma$ & 88.54$\pm$9.78 & 88.50$\pm$9.55 & 89.12$\pm$9.02 & 89.20$\pm$9.24 & 89.24$\pm$9.12 & 89.08$\pm$9.21 & \textbf{89.35$\pm$8.97} \\
\bottomrule
\end{tabular}
\end{table*}

\begin{table*}[htbp]
\centering
\caption{Single-shot readout accuracy (\%) at 512\,ns. Conventions as in Table~\ref{tab:acc_8q_704ns}; $^{\ast}$accuracy-only, never synthesized.}
\label{tab:acc_8q_512ns}
\setlength{\tabcolsep}{4pt}
\begin{tabular}{lccccccc}
\toprule
Qubit & GMM~\cite{jeffrey2014fast}$^{\ast}$ & QubiCML~\cite{vora2026ml} & MF-NN~\cite{mude2025efficient} & FCN$^{\ast}$~\cite{lienhard2022deep} & LSTM~\cite{jung2026timeseries} & LSTM-heavy & LSTM-compact \\
\midrule
Q0 & 98.04$\pm$0.11 & 97.98$\pm$0.12 & 97.82$\pm$0.12 & 98.09$\pm$0.11 & 98.17$\pm$0.11 & 97.96$\pm$0.12 & \textbf{98.17$\pm$0.11} \\
Q1 & 97.70$\pm$0.12 & 97.73$\pm$0.12 & 97.75$\pm$0.12 & 97.73$\pm$0.12 & \textbf{98.05$\pm$0.11} & 97.49$\pm$0.13 & 97.99$\pm$0.12 \\
Q2 & 97.35$\pm$0.14 & 97.30$\pm$0.14 & 97.31$\pm$0.14 & 97.21$\pm$0.14 & \textbf{97.51$\pm$0.13} & 97.08$\pm$0.14 & 97.50$\pm$0.13 \\
Q3 & 92.53$\pm$0.22 & 94.18$\pm$0.20 & 95.72$\pm$0.17 & 96.15$\pm$0.16 & 95.74$\pm$0.17 & \textbf{96.37$\pm$0.16} & 95.90$\pm$0.17 \\
Q4 & 89.03$\pm$0.26 & 88.78$\pm$0.26 & \textbf{91.28$\pm$0.23} & 90.57$\pm$0.24 & 91.26$\pm$0.24 & 90.45$\pm$0.24 & 91.27$\pm$0.24 \\
Q5 & 97.49$\pm$0.13 & 97.35$\pm$0.13 & 97.49$\pm$0.13 & 97.60$\pm$0.13 & 97.75$\pm$0.12 & 97.45$\pm$0.13 & \textbf{97.76$\pm$0.12} \\
Q6 & 92.48$\pm$0.22 & 92.76$\pm$0.21 & 92.50$\pm$0.22 & 92.94$\pm$0.21 & 93.08$\pm$0.21 & 92.94$\pm$0.21 & \textbf{93.14$\pm$0.21} \\
Q7 & 94.81$\pm$0.18 & 94.89$\pm$0.18 & 94.64$\pm$0.18 & 94.94$\pm$0.18 & 95.10$\pm$0.18 & 94.94$\pm$0.18 & \textbf{95.13$\pm$0.18} \\
\midrule
$\mu \pm \sigma$ & 94.93$\pm$3.09 & 95.12$\pm$2.99 & 95.56$\pm$2.38 & 95.65$\pm$2.51 & 95.83$\pm$2.39 & 95.59$\pm$2.48 & \textbf{95.86$\pm$2.38} \\
\bottomrule
\end{tabular}
\end{table*}

\begin{table*}[htbp]
\centering
\caption{Single-shot readout accuracy (\%) at 768\,ns. Conventions as in Table~\ref{tab:acc_8q_704ns}; $^{\ast}$accuracy-only, never synthesized.}
\label{tab:acc_8q_768ns}
\setlength{\tabcolsep}{4pt}
\begin{tabular}{lccccccc}
\toprule
Qubit & GMM~\cite{jeffrey2014fast}$^{\ast}$ & QubiCML~\cite{vora2026ml} & MF-NN~\cite{mude2025efficient} & FCN$^{\ast}$~\cite{lienhard2022deep} & LSTM~\cite{jung2026timeseries} & LSTM-heavy & LSTM-compact \\
\midrule
Q0 & 97.58$\pm$0.13 & 97.68$\pm$0.12 & 97.97$\pm$0.12 & 98.07$\pm$0.11 & \textbf{98.18$\pm$0.11} & 97.93$\pm$0.12 & 98.16$\pm$0.11 \\
Q1 & 98.52$\pm$0.10 & 98.53$\pm$0.10 & 98.50$\pm$0.10 & 98.59$\pm$0.10 & 98.75$\pm$0.09 & 98.24$\pm$0.11 & \textbf{98.75$\pm$0.09} \\
Q2 & 97.52$\pm$0.13 & 97.71$\pm$0.13 & 97.74$\pm$0.13 & 97.63$\pm$0.13 & 97.92$\pm$0.12 & 97.53$\pm$0.13 & \textbf{97.98$\pm$0.12} \\
Q3 & 88.99$\pm$0.26 & 93.92$\pm$0.20 & 95.88$\pm$0.17 & 96.39$\pm$0.16 & 95.88$\pm$0.17 & \textbf{96.60$\pm$0.15} & 96.14$\pm$0.16 \\
Q4 & 95.24$\pm$0.18 & 95.03$\pm$0.18 & 96.12$\pm$0.16 & 95.82$\pm$0.17 & 96.11$\pm$0.16 & 95.46$\pm$0.17 & \textbf{96.16$\pm$0.16} \\
Q5 & 97.85$\pm$0.12 & 97.92$\pm$0.12 & 97.87$\pm$0.12 & 97.99$\pm$0.12 & 98.20$\pm$0.11 & 97.87$\pm$0.12 & \textbf{98.25$\pm$0.11} \\
Q6 & 92.88$\pm$0.21 & 93.46$\pm$0.20 & 92.83$\pm$0.21 & 93.84$\pm$0.20 & 94.18$\pm$0.19 & 94.04$\pm$0.19 & \textbf{94.25$\pm$0.19} \\
Q7 & 94.71$\pm$0.18 & 94.70$\pm$0.18 & 94.63$\pm$0.18 & 94.90$\pm$0.18 & \textbf{95.14$\pm$0.18} & 94.73$\pm$0.18 & 95.07$\pm$0.18 \\
\midrule
$\mu \pm \sigma$ & 95.41$\pm$3.02 & 96.12$\pm$1.91 & 96.44$\pm$1.84 & 96.65$\pm$1.59 & 96.80$\pm$1.58 & 96.55$\pm$1.51 & \textbf{96.84$\pm$1.56} \\
\bottomrule
\end{tabular}
\end{table*}

\end{document}